\documentclass[aps,twocolumn,secnumarabic,balancelastpage,amsmath,amssymb,nofootinbib,floatfix,groupedaddress,superscriptaddress]{revtex4-1}

\usepackage{float}
\usepackage{graphicx}      
 
\usepackage{subcaption}

\usepackage{lgrind}        
\usepackage{bm}            

\usepackage{braket}
\usepackage{thumbpdf}
\usepackage[colorlinks = true,
            linkcolor = blue,
            urlcolor  = blue,
            citecolor = blue,
            anchorcolor = blue]{hyperref}  

\newcommand{\tripletFTwo}{$^3F_2^o$}
\newcommand{\tripletDJ}[1]{$^3D_{#1}$}
\newcommand{\singletDTwo}{$^1D_2$}
\newcommand{\singletSZero}{$^1S_0$}
\newcommand{\singletPOne}{$^1P_1^o$}
\newcommand{\tripletPOne}{$^3P_1^o$}

\begin{document}
\title{Spectroscopic Study and Lifetime Measurement of the $6d7p$  $ ^{3}F_{2}^{o}$ state of radium}

\author{D. W. Booth}
\affiliation{Physics Division, Argonne National Laboratory, Argonne, Illinois 60439, USA}
\author{T. Rabga}
\affiliation{Physics Division, Argonne National Laboratory, Argonne, Illinois 60439, USA}
\affiliation{National Superconducting Cyclotron Laboratory and Department of Physics and Astronomy, Michigan State University, East Lansing, Michigan 48824, USA}
\author{R. Ready}
\affiliation{National Superconducting Cyclotron Laboratory and Department of Physics and Astronomy, Michigan State University, East Lansing, Michigan 48824, USA}
\author{K. G. Bailey}
\affiliation{Physics Division, Argonne National Laboratory, Argonne, Illinois 60439, USA}
\author{M. Bishof}
\affiliation{Physics Division, Argonne National Laboratory, Argonne, Illinois 60439, USA}
\author{M. R. Dietrich}
\affiliation{Physics Division, Argonne National Laboratory, Argonne, Illinois 60439, USA}
\author{J. P. Greene}
\affiliation{Physics Division, Argonne National Laboratory, Argonne, Illinois 60439, USA}
\author{P. Mueller}
\affiliation{Physics Division, Argonne National Laboratory, Argonne, Illinois 60439, USA}
\author{T. P. O'Connor}
\affiliation{Physics Division, Argonne National Laboratory, Argonne, Illinois 60439, USA}
\author{J. T. Singh}
\affiliation{National Superconducting Cyclotron Laboratory and Department of Physics and Astronomy, Michigan State University, East Lansing, Michigan 48824, USA}

\date{\today}

\begin{abstract}
We report a method for the precision measurement of the oscillator strengths and the branching ratios of the decay channels of the $6d7p$  \tripletFTwo{} state in $^{226}$Ra. This method exploits a set of metastable states present in Ra, allowing a measurement of the oscillator strengths that does not require knowledge of the number of atoms in the atomic beam. We measure the oscillator strengths and the branching ratios for decays to the $7s6d$  \tripletDJ{1}, $7s6d$  \tripletDJ{2}, and $7s6d$  \singletDTwo{} states and constrain the branching ratio to the $7s6d$  \tripletDJ{3} state to be less than 0.4$\%$ (68$\%$ confidence limit). The lifetime of the \tripletFTwo{} state is determined to be $15 \pm 4$ ns. 
\end{abstract}

\maketitle

\section{Introduction}
A non-zero permanent electric dipole moment (EDM) in a non-degenerate system would indicate a new, clean signature of charge-parity (\textit{CP}) violation and contribute to our current understanding of the observed baryon asymmetry in the Universe. Octupole deformation and the nearly degenerate parity doublet in the ground state of the $^{225}$Ra atom enhances its intrinsic nuclear Schiff moment, making it an ideal candidate for an atomic EDM search~\cite{auerbach96,spevak97,dzuba02,engel05,singh15}. The best limit on the atomic EDM of $^{225}$Ra comes from the cold atom experiment at the Argonne National Laboratory~\cite{mike16}. 

\begin{figure}[h]
    \centering
    \begin{subfigure}[b]{0.22\textwidth}
        \includegraphics[width=\textwidth]{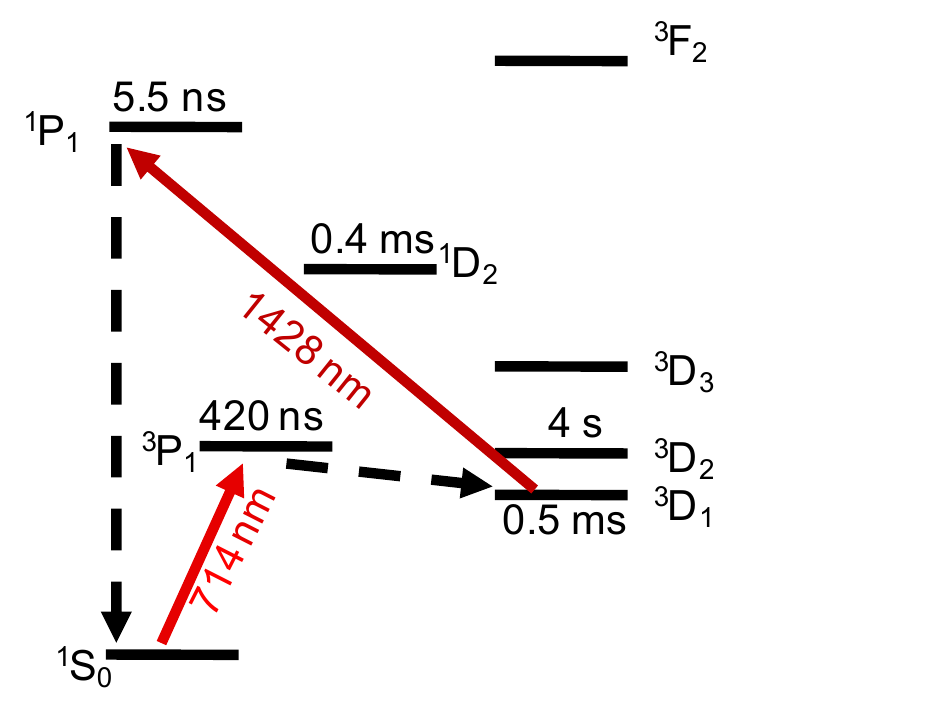}
        \caption{red slower.}
        \label{fig:redslower}
    \end{subfigure}
    ~ 
    \begin{subfigure}[b]{0.22\textwidth}
        \includegraphics[width=\textwidth]{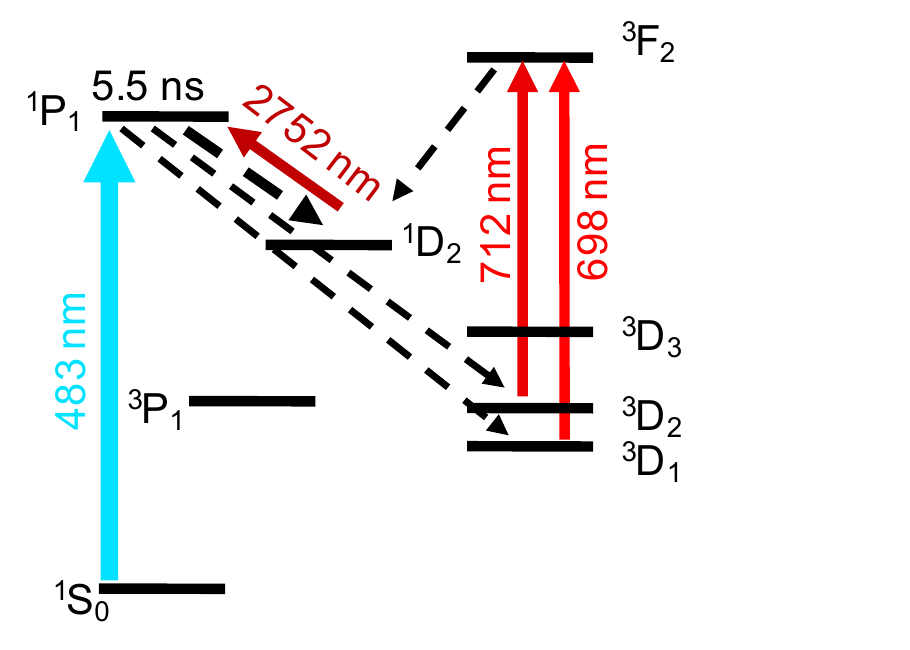}
        \caption{blue slower.}
        \label{fig:blueslower}
    \end{subfigure}
    \caption{Relevant energy levels of neutral radium along with the current and proposed improved atom slowing schemes are shown. (a) Current slowing scheme: red slower, uses the 714 nm, inter-combination line along with one repump laser at 1428 nm to longitudinally cool and slow the atoms. (b) Proposed improved slowing scheme: blue slower, will use the stronger dipole allowed 483 nm transition. The \tripletFTwo{} state has a theoretically determined lifetime of 33 ns. It will be used as a spin flipping channel to transfer atoms from the \tripletDJ{J} states into the \singletDTwo{} state, from where they are excited to the \singletPOne{} state, returning them into the cooling cycle.}
    \label{fig:slowerscheme}
\end{figure}

As a first step in trapping Ra, we must first slow the longitudinal momentum of the atomic beam. Currently, with a 1 m long Zeeman slower, using the $7s^{2}$  \singletSZero{} $\leftrightarrow$ $7s7p$  \tripletPOne{} transition at 714 nm (``red slower" scheme), as shown in Fig.~\ref{fig:redslower}, we slow atoms moving at speeds up to 60 m/s. The proposed upgrade to the cooling scheme will use the stronger dipole allowed $7s^{2}$  \singletSZero{} $\leftrightarrow$ $7s7p$  \singletPOne{} transition at 483 nm  (``blue slower" scheme), enabling us to trap atoms with speeds up to 310 m/s, resulting in an estimated two orders of magnitude more atoms. This would improve the statistical sensitivity of our EDM measurement by an order of magnitude. The improved slowing scheme is shown in Fig.~\ref{fig:blueslower}.  This slowing scheme, along with other upgrades underway, will enable the $^{225}$Ra EDM limit to set one of the most stringent constraints on the hadronic $CP$ violating parameters~\cite{chupp19}.

As shown in Fig.~\ref{fig:redslower}, we currently cool and trap the atoms using the $7s^{2}$  \singletSZero{} $\leftrightarrow$ $7s7p$  \tripletPOne{} inter-combination line instead of the stronger dipole allowed $7s^{2}$  \singletSZero{} $\leftrightarrow$ $7s7p$  \singletPOne{} transition. This has the advantage of requiring only one `repump' laser at 1428 nm to depopulate the \tripletDJ{1} state back into the cooling cycle. 

The blue slower upgrade to our atom slowing scheme will use the stronger dipole allowed transition, as shown in Fig.~\ref{fig:blueslower}. However, the atoms excited to the \singletPOne{} state decay to the long-lived \singletDTwo{}, \tripletDJ{1}, and \tripletDJ{2} states. This requires additional lasers to depopulate these states. A na\"{i}ve repumping scheme would transfer the atoms from the various \textit{D}-states into the \singletPOne{} state, returning them into the cooling cycle. This presents several challenges. The additional repump transitions have significantly different magnetic moments, making it impossible to design a magnetic field profile that simultaneously compensates for the associated Doppler shifts for a Zeeman slower. In addition, given the weak transition strengths from the \tripletDJ{J} manifold to the \singletPOne{} state, we would require higher laser powers than are readily available. 

Theoretical studies of the low lying atomic levels of Radium suggests strong spin-orbit interaction in the $6d7p$  \tripletFTwo{} state~\cite{flambaum06}. The E1-transition amplitude for the spin forbidden \tripletFTwo{} $\rightarrow$ \singletDTwo{} transition is enhanced, while the spin allowed \tripletFTwo{} $\rightarrow$ \tripletDJ{3} transition is greatly suppressed. These properties are consistent with a \textit{j-j} coupled state having the term symbol $(6d_{3/2} 7p_{1/2})$. The proposed blue slower scheme will employ this strong spin-orbit coupling in the \tripletFTwo{} state, using it as a spin flipping channel to transfer atoms from the \tripletDJ{J} states into the \singletDTwo{} state, from which they can be excited to the \singletPOne{} state, returning them to the cooling cycle. 

Of the four transitions involved, the  \tripletFTwo{} $\rightarrow$ \singletDTwo{} and the \tripletFTwo{} $\rightarrow$ \tripletDJ{3} transitions have not been experimentally observed before~\cite{Dammalapati16,Ramussen34}. This necessitated a spectroscopic study of the \tripletFTwo{} state, which involved measuring the oscillator strengths of the transitions to this state from the relevant \textit{D}-states and the respective branching ratios.  

Precise measurements of oscillator strengths and the branching ratios provide effective tests of atomic structure calculations, especially where electron correlation effects~\cite{hibbert75} and relativistic corrections~\cite{hibbert77} are important. They are also important for atomic parity non-conservation searches in the unified electro-weak theories~\cite{huber86}. Oscillator strengths can be determined either directly or indirectly. Direct measurements rely on absolute measurements of absorption, emission, or dispersion of relevant transitions, while indirect methods include combined measurements of the branching ratios and the lifetimes of the relevant higher states. Direct measurements require the knowledge of the sample number density as well as local thermodynamic equilibrium~\cite{huber86}. However, since the \tripletFTwo{} state has large branching ratios to several long-lived states, it is suitable for a novel method for measuring the oscillator strengths that circumvent these requirements. This method is described below.

\section{Method}

The method that we employ for measuring the oscillator strengths of the transitions from the different \textit{D}-states to the \tripletFTwo{} state requires that when `probed' on a certain \textit{D}-state to \tripletFTwo{} state transition, the atoms scatter only a few photons from the probe laser before they decay into a meta-stable dark state. The lifetimes of the \singletDTwo{} and \tripletDJ{1} states are measured to be 385(45) $\mu$s~\cite{trimble09} and 510(60) $\mu$s~\cite{guest07} respectively, and the calculated lifetime of the \tripletDJ{2} state is determined to be 4 s~\cite{Bieron04}. By comparison, the average transit time of the atoms through the beam is 38 $\mu$s.

For the \tripletDJ{3} we set an upper bound on the branching ratio by comparing the fluorescence signal as the atom decays into this state, compared to the fluorescence signal as it decays into the \singletDTwo{} state. For the rest, we selectively `probe' the atoms in the three different states, \singletDTwo{}, \tripletDJ{2} and \tripletDJ{1} by exciting them into the \tripletFTwo{} state and observing the fluorescence as they decay to one of the other non-probed states. Since we use an atomic beam for our measurements, an atom that decays into an unprobed or ``dark", long-lived \textit{D}-state, is not re-excited by the probe beam. For instance, the \tripletFTwo{} state decays with a maximum branching ratio of about 60$\%$ into the \tripletDJ{2} state, which corresponds to on average 2.5 photons/atom scattered on this transition before going dark. The probability with which the atom scatters a probe photon is a Poisson process dependent on the oscillator strength of the transition, as well as the natural linewidth of the \tripletFTwo{} state, which is proportional to the sum of the oscillator strengths of all its possible decay channels. An atom that decays from the \tripletFTwo{} state to the probe state with a probability \textit{p}, has a probability of $p^{n-1}(1-p)$ of decaying into a dark state after \textit{n} scattering events. Here \textit{p} is the branching ratio from the \tripletFTwo{} state to the probe state and is a function of the oscillator strengths. Thus, the overall process is described by a Poisson and geometric distribution. If we probe all the relevant transitions to the excited state of interest, we can fit for every oscillator strength. We also need to know the oven temperature, although it will be shown later that this does not need to be very precise. It also requires a measurement of the probe beam profile and the probe beam powers. Apart from these, this method circumvents the need for atomic flux calibration and the absolute efficiency of various parts of the experimental setup. As for the \tripletFTwo{} state, this provides an effective technique for oscillator strength measurements in an atomic beam, where one probes a state directly inaccessible from the ground state, and with significant branching ratios to several lower lying, long-lived states. 

\section{Experimental Setup}
\begin{figure}[h]
\includegraphics[width=0.45\textwidth]{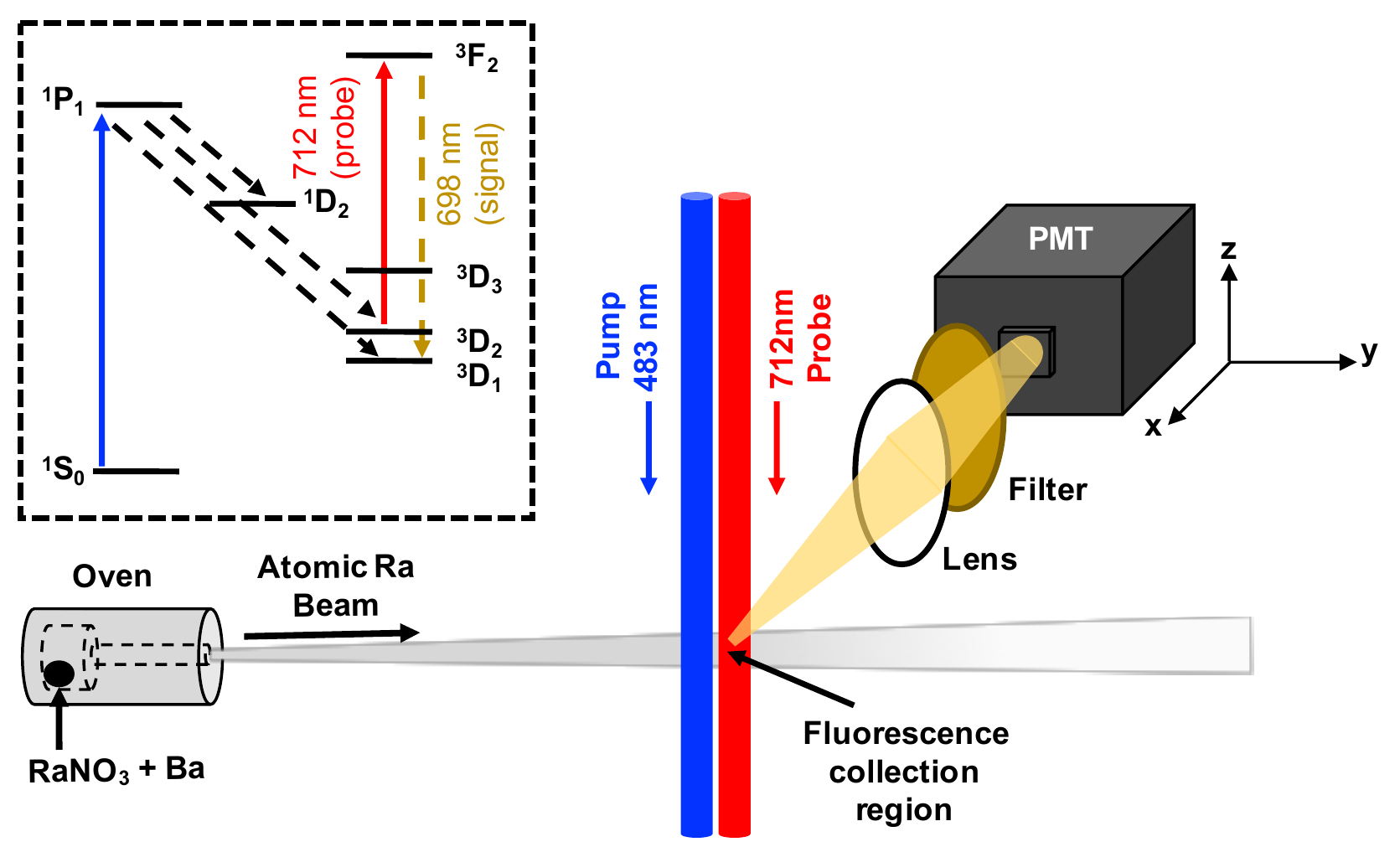}
\caption{Spectroscopy measurement setup (not to scale). The atoms exiting the oven are pumped into the \singletPOne{} state from which they decay into the \textit{D}-states. Here, we probe the atoms in the \tripletDJ{2} state with a resonant 712 nm probe laser beam which excites them into the \tripletFTwo{} state and the resultant fluorescence 698 nm is filtered and detected on a PMT. Inset: the relevant energy levels and the transitions involved for probing the atoms in the \tripletDJ{2} state and detecting the fluorescence as the atoms decay from \tripletFTwo{} $\rightarrow$ \tripletDJ{1}. }
\label{fig:setup}
\end{figure}

The measurements presented in this work were performed on $^{226}$Ra ($t_{1/2} = 1600\,\text{y}$, I = 0). The experimental setup is shown in Fig.~\ref{fig:setup}. We load 3 $\mu$Ci of $^{226}$Ra in the form of radium nitrate salt along with two 25 mg pieces of metallic barium into the oven and radiatively heat it. The barium helps chemically separate the radium from the radium salt, allowing a steady flux of atomic radium out of the nozzle. The nozzle geometry limits the divergence of the atomic beam to 50 mrad \cite{mike16}. About 30 cm from the nozzle, the atoms are excited into the \singletPOne{} state using 3 mW of 483 nm `pump' light produced by a Toptica DL Pro diode laser which is frequency stabilized to a zerodur opitical cavity. The atoms then decay to the long-lived \textit{D}-states as shown in the inset of Fig.~\ref{fig:setup}. The probe laser beams at 912 nm, 712 nm and 698 nm for respectively probing the \singletDTwo{}, \tripletDJ{2}, and \tripletDJ{1} state are produced by a Sirah Matisse Ti:Sapphire ring cavity laser which is frequency stabilized to a high finesse ULE optical cavity. For instance, in Fig.~\ref{fig:setup}, we probe the atoms in the \tripletDJ{2} state by resonantly exciting the atoms into the \tripletFTwo{} state with a 712 nm probe beam and observing the subsequent decay of the atoms into the \tripletDJ{1} state at 698 nm. We collect line-shape data for a given probe transition and laser power by scanning the probe laser frequency. This is repeated for a range of different probe powers. 

We scan the frequency of the probe laser by double passing it through an acousto-optic modulator (AOM) and scanning its input RF frequency. The fluorescence signal is collected by a 2'' diameter, 10 cm focal length lens, filtered and focused onto a photo-multiplier tube (PMT) (Hamamatsu, H7421-50). The PMT output is sent to a data acquisition (DAQ) card (NI- USB6341). The pump laser is shuttered a rate of 0.5 Hz to subtract any scattered background from the probe light on the PMT in the fluorescence signal. This is repeated for all the probe transitions of interest. The probe beam intensity profile is measured using a Thorlabs DCC1545M CMOS camera ($1280 \times 1024$ pixels).

To limit the branching ratio of the \tripletFTwo{} $\rightarrow$ \tripletDJ{3} transition, with the probe laser at 912 nm, probing the \singletDTwo{} $\rightarrow$ \tripletFTwo{} transition, we compare the signal at 750 nm (decay to the \tripletDJ{3} state), with the signal strength at 698 nm (decay to the \tripletDJ{1} state) using narrow bandwidth filters. Using the value we obtain for the 698 nm branching ratio from our analysis, we convert the relative limit on the branching ratio of \tripletDJ{1} $\rightarrow$ \tripletFTwo{} transition to an absolute limit.

\begin{figure}[t]
    \centering
    \includegraphics[width=0.5\textwidth]{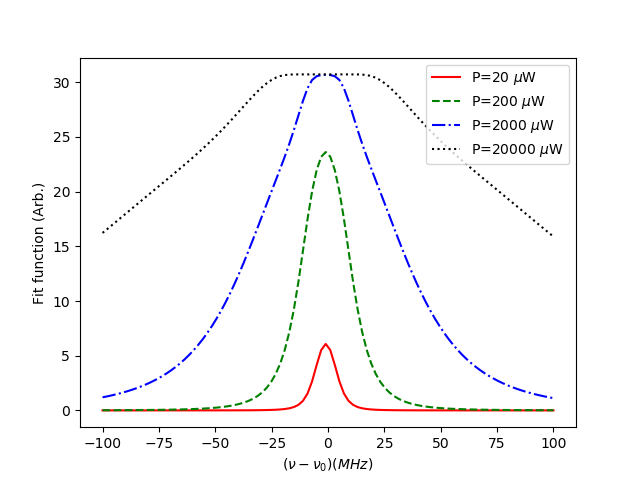}
    \caption{An example of the behavior of the fit function as the probe beam power is increased. This example is shown for the 712 nm probe transition, which has an oscillator strength $f_{ik}=0.32(12)$. }
    \label{fig:samplefits}
\end{figure}

\begin{figure*}[t]
    \centering
    \includegraphics[width=1\textwidth]{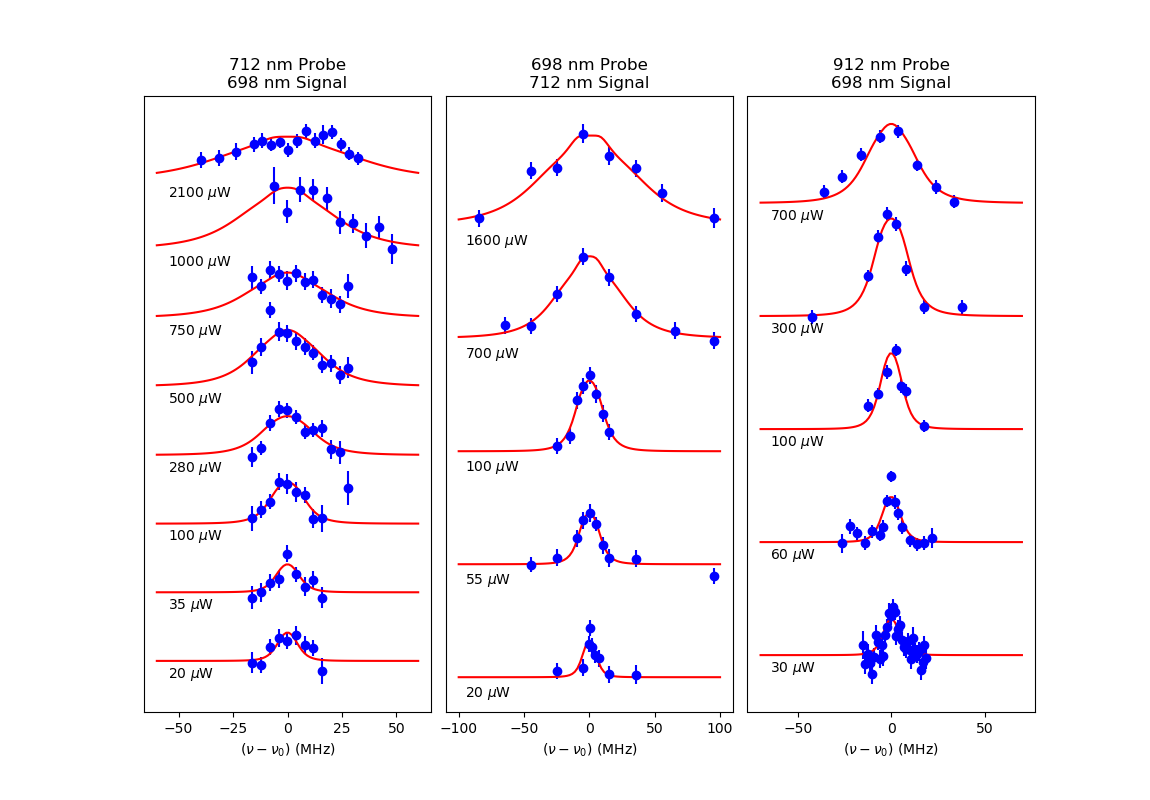}
    \caption{Waterfall plot of the data and the fits for each of the lineshapes measured in the experiment. The count rate for a particular probe transition at a given power is plotted against the detuning ($\nu-\nu_{0}$) from resonance. The lineshapes are offset vertically for clarity. }
    \label{fig:fits}
\end{figure*}

\section{Data Analysis}

We collect lineshape data for 8 different powers with the probe at 712 nm and 5 different powers each with the probe at 698 nm and 912 nm, a total of 18 lineshape data sets.
For each of the lineshape data sets measured, we fit the fluorescence counts at the signal transition to a function `$F$' to extract the oscillator strengths:

\begin{equation}
    F = A \sum_{n=1}^{n_{\mathrm{max}}} p_{ik}^{n-1} (1-p_{ik}) \left[1-\mathrm{CDF}(\Lambda,n)\right] + C_0 ,
    \label{eqn:fitfn}
\end{equation}

\noindent where $n$ is the number of photons scattered before the atom decays into the signal state, $p_{ik}$ is the branching ratio from the \tripletFTwo{} state to the probe state $k$, $A$ is an amplitude extracted from the fit, and $C_0$ is a constant offset. The nominally infinite sum is truncated to $n_{\mathrm{max}}$ and checked for convergence. The term $p_{ik}^{n-1}(1-p_{ik})$ is the geometric probability that the atom decays into a dark state after on the $n^\mathrm{th}$ scattering event from the probe state $k$.

The cumulative distribution function, $\mathrm{CDF}$, for a Poissonian distribution that describes our probing process, is the probability that up to $n$ probe photons are scattered. The term $1-\mathrm{CDF}$ represents the probability that the probe laser power is sufficient to scatter at least \textit{n} photons, and the cumulative term $p_{ik}^{n-1}(1-p_{ik})[1-\mathrm{CDF}]$ gives the probability that after scattering \textit{n} probe photons the atom decays into our signal state. The $\mathrm{CDF}$ is given by

\begin{equation}
    \mathrm{CDF}(\Lambda,n) = \frac{\Gamma[\mathrm{floor}(n+1),\Lambda]}{\mathrm{floor}(n)!} ,
\end{equation}

\noindent where $\Gamma$ is the upper incomplete gamma function. The beam shape is incorporated into the expression for the Poissonian weight, $\Lambda$, given by:

\begin{equation}
    \Lambda = \sum_{y} \tau \bar{n}(y) f \sigma_0 g(\nu-\nu_0;\Gamma,\gamma_{D}) ,
    \label{eqn:lambda}
\end{equation}

\noindent where $\tau$ is the interaction time as the atom passes through one pixel of the laser beam image of width $\Delta y$, $f$ is the oscillator strength, $\nu$ is the laser frequency, $\nu_0$ is the line center, and $\sigma_0 = \pi r_e c = 2.654 \times 10^{-2} \mathrm{cm}^2/s$, where $r_e$ is the classical electron radius and $c$ is the speed of light \cite{Axner04}. $g$ is the lineshape function and $\bar{n}(y)$ is the photon intensity at a given \textit{y}-pixel. The interaction time is determined by the velocity distribution of the beam:

\begin{eqnarray}
    \tau = \Delta y\left<\frac{1}{v}\right> \\
    \left<\frac{1}{v}\right> = \int_0^\infty B(v;T) \frac{1}{v} dv ,
\end{eqnarray}

\noindent where the Boltzmann velocity distribution $B(v;T)$ is defined as \cite{MCArxiv}:

\begin{eqnarray}
    B(v;T) = 2 \frac{u^2}{\bar{v}} e^{-u^2} \\
    u = v/\bar{v} ,
\end{eqnarray}

where $\bar{v}$ is the most probable speed.

The photon intensity observed by an atom at horizontal position $y$, $\bar{n}(y)$ is:

\begin {equation}
    \bar{n}(y) = \frac{P}{h \nu \Delta y^2} \frac{\sum_{x} I(x,y)}{\sum_{x,y} I(x,y)} ,
\end{equation}

\noindent where $P$ is the total power in the laser beam, $\Delta y$ is the width on one pixel on the beam image (5.2 $\mu$m), and $I(x,y)$ is the beam strength at pixel $(x,y)$. 

The lineshape function $g(\nu-\nu_0;\Gamma,\gamma_{D})$ is a Voigt profile, with the Gaussian width $\gamma_{D}$ determined, by a measurement of the Doppler-broadened linewidth on the narrow 714 nm, \singletSZero{} $ \rightarrow\, $ \tripletPOne{} transition ($\Gamma_{714} = 2\pi \times 380\,\text{kHz}$), to be $2.32$ MHz. The Lorentzian width of the Voigt profile is a dependent parameter of the fit:

\begin{equation}
    \Gamma = \sum_{i} A_{ki} = \frac{2 \pi^2 r_e c}{2J_k+1} \sum_i \frac{2J_i+1}{\lambda^2_{ik}} f_{ik} ,
\end{equation}

\noindent where $A_{ki}$ is the spontaneous decay transition rate from an upper state $k$ to a lower state $i$, and $J_i$ and $J_k$ are the total angular momenta of states $i$ and $k$ respectively. The branching ratios $p_{ik}$ can be expressed in terms of $A_{ki}$ as:

\begin{equation}
    p_{ik} = \frac{A_{ki}}{\Gamma} .
\end{equation}

The parameters of the fit are the oscillator strengths $f_{ik}$, the peak amplitudes $A$, the peak centers $\nu_0$, and the offsets $C_0$. The offsets, peak centers, and oscillator strengths are common for fits of peaks from the same transition. The amplitudes vary between different lineshapes due to variations in the atom flux from the oven. The offsets are to account for various background sources. Examples of the effect of the oscillator strength on the fit function are shown in Fig. \ref{fig:samplefits}. As the power increases, it results in a broader peak for a fixed oscillator strength.

The lineshapes for each power taken for each transition are cumulatively fit, using a least squares method using the curve fitter from the scipy library, version 1.1.0, in Python.

\section{Results}

Fits for each of the lineshapes measured are shown in Fig.~\ref{fig:fits}. The reduced chi-squared for the cumulative fit is $\chi_{\nu}^2=1.13$, (179 DOF). Results for the oscillator strengths and branching ratios are shown in Table \ref{tab:oscstr}. From these oscillator strengths we can determine the linewidth of the \tripletFTwo{} state to be $10.4 \pm 2.7$ MHz.

Convergence of the sum in Eqn.~\ref{eqn:fitfn} is tested by varying the upper limit on $n$, testing the values $n_{\mathrm{max}}=[5,10,20,40]$. Results of this test are shown in Table \ref{tab:convergence}. For $n_{\mathrm{max}}=10$ and above, there is strong convergence for all of the oscillator strengths.



\begin{table}[h]
    \centering
    \begin{tabular}{c|c|c|c}
       $n_{\text{max}}$  & \tripletFTwo{} $\rightarrow$ \tripletDJ{2}      & \tripletFTwo{} $\rightarrow$ \tripletDJ{1}     & \tripletFTwo{} $\rightarrow$ \singletDTwo{}     \\  \hline
       1                 & 0.247(44)   & 0.234(39)  & 0.046(7)   \\
       5                 & 0.294(71)   & 0.254(65)  & 0.042(7)   \\
       10                & 0.318(105)  & 0.243(79)  & 0.041(8)   \\
       20                & 0.318(119)  & 0.245(87)  & 0.041(9)   \\
       40                & 0.317(120)  & 0.245(87)  & 0.041(9)   \\
    \end{tabular}
    \caption{Convergence of the oscillator strengths as a function of the number of terms, $n_{\text{max}}$, in the sum in Eqn.~\ref{eqn:fitfn}.}
    \label{tab:convergence}
\end{table}

\begin{table}[h]
    \centering
    \begin{tabular}{c|c|c|c}
       $T (^{\circ}\mathrm{C})$  & \tripletFTwo{} $\rightarrow$ \tripletDJ{2}      & \tripletFTwo{} $\rightarrow$ \tripletDJ{1}     & \tripletFTwo{} $\rightarrow$ \singletDTwo{}     \\  \hline
       420                       & 0.317(120)  & 0.239(85)  & 0.040(9)   \\
       470                       & 0.318(119)  & 0.245(87)  & 0.041(8)   \\
       520                       & 0.319(119)  & 0.251(88)  & 0.042(9)   \\
    \end{tabular}
    \caption{The oscillator strengths as a function of the temperature of the Radium oven, with $n_{\text{max}}=20$.}
    \label{tab:tempdependence}
\end{table}

An additional factor that may affect our results is the temperature of the radium oven source. There can be uncertainty in this temperature due to temperature gradients between inside of the oven crucible and the monitor thermocouple at the rear of the crucible. This would cause a systematic shift in the Boltzmann factor $B(v;T)$, which affects the interaction time $\tau$ of the atoms with the laser beam in Eqn.~\ref{eqn:lambda}. To test this, we fit the data using $n_{\mathrm{max}}=20$ and $T=[420, 470, 520] \mathrm{^{\circ}C}$. Results of this test are shown in Table \ref{tab:tempdependence}. We find both of these systematic effects to be negligible.

\begin{table*}[t]
\caption{Wavenumbers, Oscillator strengths ($f_{ik}$), and branching ratios (BR) for the transitions out of the \tripletFTwo{} state. Theory values for the branching ratio from \cite{flambaum06} are shown for comparison. }
\label{tab:oscstr}
\begin{tabular}{l|c|c|c|c}
Transition  & Wavenumber (cm$^{-1}$) & $f_{ik}$ (theory) & $f_{ik}$  &  BR  \\ \hline
\tripletFTwo{} $\rightarrow$ \singletDTwo{}  & 10956.7095(5)  & 0.054 &   0.041(9)  & 0.050(11) \\  
\tripletFTwo{} $\rightarrow$ \tripletDJ{2}  & 14044.0875(5) & 0.074 & 0.32(12) & 0.64(24) \\
\tripletFTwo{} $\rightarrow$ \tripletDJ{1}  & 14322.2340(5) & 0.202 & 0.25(8) & 0.31(11)
\end{tabular}
\end{table*}

A limit on the \tripletFTwo{} $\rightarrow$ \tripletDJ{3} branching ratio was determined by comparing the strength of the signal at 750 nm from this transition to the \tripletDJ{1} signal strength at 698 nm. A 105-minute integration of the 750 nm signal level resulted in $2.5 \pm 0.75$ counts per second, compared to a 10 minute integration at 698 nm, where we had a signal level of $73 \pm 2.9$ counts per second.

A source of systematic error in this measurement that had to be accounted for is the leakage of 483 nm light through the 750 nm filter. This leakage causes a constant offset on the signal level, which is accounted for in the lineshape fitting by adding a constant offset. In the 750 nm measurement, however, this error can appear as a false detection of atoms decaying on this decay channel. To determine the effect of 483 nm leakage, we measured the PMT signal in the absence of probe light for a period of 14 hours, yielding a signal level of $2.43 \pm 0.28$ counts per second. After removing this effect, the result at 750 nm is $0.08 \pm 0.81$ counts per second, which we interpret as an upper limit of $0.4\%$ (68$\%$ C.L.) on the branching ratio to \tripletDJ{3}. This upper bound is computed using the Feldman-Cousins contruction of confidence intervals for a Gaussian distribution constrained to have a non-negative mean.~\cite{Feldman98}

\section{Discussion}

Although differing in detail from the prediction in \cite{flambaum06}, the overall picture of strong spin-orbit coupling in \tripletFTwo{} is confirmed. The \tripletDJ{3} is found to be highly suppressed, and the \singletDTwo{} strongly enhanced compared to the normal dipole selection rules for a \tripletFTwo{} state. The oscillator strength of the \tripletDJ{2} channel is found to be significantly higher than predicted, although the others are in good agreement. It should be noted, however, that the \tripletFTwo{} $\rightarrow\, ^{1}D_{2}$ transition is a forbidden transition in the absence of relativistic effects with its branching ratio component coming from the strong spin-orbit interaction in the \tripletFTwo{} state.  

For the \tripletFTwo{} $\rightarrow\, $\tripletDJ{3} transition, the branching ratio from the dipole matrix elements from \cite{flambaum06} is $3.6 \times 10^{-4}$, which is an order of magnitude below our upper bound of $4 \times 10^{-3}$ (68$\%$ C.I.).

The theoretical value for the lifetime of the \tripletFTwo{} state is 33 ns \cite{flambaum06}, which differs from our experimental lifetime (derived from the natural linewidth) of $15.3 \pm 4.0$ ns, by about a factor of two. 

Our results for the branching ratios out of the \tripletFTwo{} are promising for the use of this state for our scheme for repumping the blue Zeeman slower. The branching ratio to the \singletDTwo{} state is high enough to make repumping through this state reasonable, despite being lower than the theoretical branching ratio. With a branching ratio of 0.05, 20 excitations would be needed on average to bring an atom into the \singletDTwo{} state. 



\section{Conclusions}
We have measured the oscillator strengths and the branching ratios of the decay channels of the \tripletFTwo{} state. We employ what we believe to be a novel technique that is used in an atomic beam measurement and requires that the atom excited to a higher energy level decays into several long-lived states, and therefore scatters only a few photons on the transition of interest. This technique requires measuring lineshape data for a number of powers for all the transitions involved. The cumulative fit to all the data is a function of the oscillator strengths of all the transitions to the excited state. This method avoids the need for knowledge of the atomic number density, as well as the absolute calibration of the efficiencies of the different parts of the experimental setup. 

The systematic effects due to the oven temperature uncertainty and the finite sum over the photon scattering events \textit{n} in the fit function Eq.~\ref{eqn:fitfn} up to an upper limit $n_{\text{max}}$ were studied. Both of these effects are negligible compared to the statistical uncertainties in the oscillator strengths. The main limitation on the maximum number of photons is computation time. Our implementation of the fitting algorithm took about a day to run with $n_{max}=40$ on a laptop computer with an Intel i7-7700HQ CPU. 

The strong decay channel to \singletDTwo{}, along with the strong suppression of the branching ratio to the $^3D_3$ state compared to the $^1D_2$ state supports the theoretical prediction of strong spin-orbit coupling in the \tripletFTwo{} state in radium.

The above spectroscopic study of the \tripletFTwo{} state supports our proposal of using this state as a spin flipping channel for implementing the blue slower upgrade to our atom slowing and cooling apparatus. This will improve our atom trapping efficiency by two orders of magnitude and enhance our sensitivity to the atomic EDM of $^{225}$Ra by at least a factor of 10, setting very stringent limits on the hadronic \textit{CP}-violating parameters.

\section{Acknowledgments}

This work is supported by the U.S. DOE, Office of Science, Office of
Nuclear Physics under contracts DE-AC02-06CH11357 and DE-SC0019455, and by Michigan State University. RR acknowledges support from the
United States Department of Energy Office of Science Graduate Student Research (SCGSR) Program.


\clearpage

\end{document}